\documentclass[pra,twocolumn,floatfix,superscriptaddress,longbibliography]{revtex4-1}

\usepackage[T1]{fontenc}
\usepackage{natbib}
\usepackage{graphicx}
\usepackage{indentfirst}
\usepackage{listings}
\usepackage{braket}
\usepackage{amsmath}
\usepackage{amssymb}
\usepackage{amsthm}
\usepackage[font=small,labelfont=bf]{caption}
\usepackage{float}
\usepackage{caption}
\usepackage{subcaption}
\usepackage{array, makecell}
\usepackage{bbold}
\usepackage{placeins}

\usepackage[title]{appendix}
\newtheorem{definition}{Definition}
\newtheorem{theorem}{Theorem}

\usepackage{qcircuit}
\usepackage{xcolor}

\usepackage{tikz}
\usepackage{lipsum}

\begin{document}

\title{Quantum representation of finite groups}

\author{Ruge Lin}
\affiliation{Quantum Research Centre, Technology Innovation Institute,  UAE.}
\affiliation{Departament de F\'isica Qu\`antica i Astrof\'isica and Institut de Ci\`encies del Cosmos, Universitat de Barcelona, Spain.}

\begin{abstract}

The concept of quantum representation of finite groups (QRFG) has been a fundamental aspect of quantum computing for quite some time, playing a role in every corner, from elementary quantum logic gates to the famous Shor's and Grover's algorithms. In this article, we provide a formal definition of this concept using both group theory and differential geometry. Our work proves the existence of a quantum representation for any finite group and outlines two methods for translating each generator of the group into a quantum circuit, utilizing gate decomposition of unitary matrices and variational quantum algorithms. Additionally, we provide numerical simulations of an explicit example on an open-access platform. Finally, we demonstrate the usefulness and potential of QRFG by showing its role in the implementation of some quantum algorithms and quantum finite automata.

\end{abstract}

\maketitle

\section{Introduction}

As stated by Zee \cite{zee2016group}, the unitarity theorem asserts that all finite groups possess unitary representations. This concept holds significant importance in the realm of quantum mechanics because the theory relies on unitary representations of symmetries within physical systems \cite{gudder2006quantum,kornyak2011computations,majid2000foundations,cornwell1997group}. Consequently, it is natural to explore the quantum representation of any finite group, where each element is redefined as an effective quantum operator. By encoding the mathematical rules governing the group into a series of physical motions, quantum representation elucidates the connection between abstract mathematical concepts and applicable quantum circuits. This framework finds particular relevance within the context of quantum computation, where the quantum representation serves as the mapping from a given finite group to a gate set, enabling its utilization in quantum algorithms.

\section{Preliminary}

In this section, we lay the foundation by introducing key definitions that establish a connection between group actions on sets and quantum operations on states. To ensure clarity and avoid confusion, we adopt a notation where the symbol "$\cdot$" represents group action, while the symbol "$\times$" denotes the multiplication of numbers. This distinction in notation facilitates a clear delineation between the two concepts, enabling a seamless transition from group theory to quantum mechanics.

\begin{definition}{Unitary action}

The unitary action $\alpha: U\left(d\right) \cdot H_d \rightarrow H_d$ is the group action of the unitary group $U\left(d\right)$ on the Hilbert space $H_d$.
\end{definition}

\begin{definition}{Ineffective quantum group}

For $x$ and $y$ elements of $H_d$, if they are indistinguishable by any measurement after any unitary action, we say that they are equivalent in the context of quantum mechanics and note $x\sim y$. In particular, the quotient space $H_d/\sim$ is called projective Hilbert space $\mathbb{P}\left(H_d\right)$. A unitary operator is ineffective if it produces no observable difference when acting on any element of $H_d$. All ineffective quantum operators form the ineffective quantum group $U\left(1\right)=\exp\left(i\theta\right)I_d$, the center of $U\left(d\right)$, where $I_d$ is the $d\times d$ identity matrix.
\end{definition}

\begin{definition}{Effective quantum group}

The effective quantum group is the quotient $U\left(d\right)/U\left(1\right)$, the projective unitary group $\mathbb{P}U\left(d\right)$. 
\end{definition}

A pure quantum state can be described by a normalized $d$-dimensional complex vector $\mathbf{v}$. Then $H_d$ is the hypersphere $\mathbb{S}^{2d-1}$, where each point of this hypersphere describes a quantum state, which is not unique because multiple points can describe the same state. Unitary action $\alpha$ is the differentiable action of the compact Lie group $U\left(d\right)$ on the smooth manifold $\mathbb{S}^{2d-1}$. For an element $M_d \in U\left(d\right)$ and a vector $\mathbf{v}$, where the component of $\mathbf{v}$ describes a point located on $\mathbb{S}^{2d-1}$, $\alpha$ is the matrix multiplication $M_d \cdot \mathbf{v} \mapsto \mathbf{v}$. This action is transitive because we can always find a unitary matrix that transfers one quantum state to another. However, the action is not free because if a unitary transfers a quantum state to itself, it does not imply this unitary is the identity. 

\begin{theorem}{Quantum state manifold}

The projective Hilbert space $\mathbb{P}\left(H_d\right)$ can be described by the quantum state manifold, defined as $\mathbb{S}^{2d-1}/U\left(1\right)$, which is a submanifold of $\mathbb{S}^{2d-2}$. Every quantum state uniquely corresponds to one point in the manifold.

\begin{proof}

The action of $U\left(1\right)$ on $\mathbb{S}^{2d-1}$ is a continuous compact Lie group action on a smooth manifold. Therefore, the action is smooth and proper. For all points on $\mathbb{S}^{2d-1}$, if one element of $U\left(1\right)$ keeps the point fixed after the action, that means $\theta=0$, and the element is the identity. So the action is also free. We have $U\left(1\right)$ Lie group acting smoothly, freely, and properly on a smooth manifold $\mathbb{S}^{2d-1}$, which allows us to apply the Quotient Manifold Theorem \cite{lee2012smooth}. Therefore $\mathbb{S}^{2d-1}/U\left(1\right)$ is a manifold of dimension $2d-2=\dim{\mathbb{S}^{2d-1}}-\dim{U\left(1\right)}$. 

The same result can be obtained when $\mathbb{P}\left(H_d\right)$ is described as a complex projective space and quantum states are its rays \cite{ashtekar1999geometrical}. Here, we reprove this to ensure that there is no ambiguity in the notation.

When $d=2$, if the coefficient of $\ket{0}$ is forced to be real and non-negative, $\mathbb{S}^{3}/U\left(1\right)=\mathbb{S}^{2}$ and we recover the representation of the Bloch sphere. For $d>2$, if the coefficient of $\ket{0...0}$ is forced to be real and non-negative, we can obtain $\mathbb{S}^{2d-1}/U\left(1\right)\subset \mathbb{S}^{2d-2}$, which means $\mathbb{P}\left(H_d\right)$ form a submanifold of hypersphere $\mathbb{S}^{2d-2}$ of the same dimension. But they are not equal since multiple points of $\mathbb{S}^{2d-2}$ can still define the same quantum state.
\end{proof}
\end{theorem}

\section{Quantum representation}

Once we have translated the concepts of "probability of all possibilities sum up to one" and "global phase is undetectable" into the language of group theory and differential geometry, although in a non-unique and non-original manner, we can proceed to provide a formal definition of the quantum representation of a group.

\begin{definition}{Quantum representation}

Given a group $G$, its quantum representation is a subset of $\mathbb{P}U\left(d\right)$ that maintains the group structure of $G$. Compared to projective unitary representations \cite{gannon2007moonshine}, we add a constraint that quantum representation should be faithful: each element in $G$ corresponds to a distinct element in $\mathbb{P}U\left(d\right)$. In particular, the identity element $e$ is always represented by the identity operator $I_d$.
\end{definition}

This extra constraint of faithfulness is crucial and implies a key difference between this article and all previous works. Not all groups have their quantum representations. But in the following, we can prove that when $G$ is finite, it always does.

\begin{theorem}{Quantum representation of finite groups}\label{th}

For a given finite group $G$, there is always a Hibert space with a certain dimension $d$ in which $G$ has its quantum representation (QRFG). Additionally, there exists $x\in \mathbb{P}\left(H_d\right)$ such that $G\cdot x$ the orbit of $x$ has exactly $|G|$ elements. This orbit forms a principal homogeneous space of $G$, denoted as the $G$-torsor.

\begin{proof}
Cayley's theorem states that every finite group $G$ is isomorphic to a subgroup of a symmetric group $S_n$, which is a subgroup of $S_d$ for $d\geq n$. We have $S_d$ subgroup of $\mathbb{P}U\left(d\right)$ since they can be represented as permutation matrices, which are orthogonal; more importantly, none of them has a global phase difference from another. The quantum representation of $G$ is a set of $|G|$ permutation matrices of size $d\times d$ that preserves the structure of $G$. In particular, when we select $x=\sum_{i=0}^{d-1}c_i\ket{i}$ with distinct $c_i$, the orbit $G\cdot x$ forms a $G$-torsor. This is a trivial mapping, but it guarantees the existence of quantum representation for every finite group.

We then prove the existence of a principal homogeneous space for every possible mapping. Given a set of unitary $\{I_d,U_{i=2},...,U_{i=|G|}\}$ which forms a quantum representation of $G$, there is always $x\in\mathbb{P}\left(H_d\right)$ such that $|G\cdot x|=|G|$. The orbit-stabilizer theorem for finite groups gives $|G\cdot x|\times|G_x|=|G|$, where $|G_x|$ is the stabilizer of $x$. For $G\cdot x$ to be the $G$-torsor, we need $G_x={I_d}$, which means $x$ should locate outside the union of all $F_i$, space of fixed point of $U_i$, defined as $F_i=\{\forall \ket{\psi}\in\mathbb{P}\left(H_d\right): U_i\ket{\psi}=\ket{\psi}\}$. For $U_i\neq I_d$, we have $\dim{F_i} < \dim{\mathbb{P}\left(H_d\right)}=2d-2$ and it is impossible to cover a space with a finite number of subspaces of lower dimension, a point $x$ always exists outside this union.
\end{proof}
\end{theorem}

We use a small example to highlight the importance of faithfulness in the definition. The Pauli group is a $16$ elements matrix group isomorphic to $C_4\circ D_4$, the central product of the cyclic group of order $4$ and the dihedral group of order $8$. However, its quantum representation only has $4$ elements and is isomorphic to $C_2\times C_2$ ($\times$ for the direct product), which is abelian. Pauli matrices are anti-commute, but quantum Pauli gates commute to a global phase. Instead of different matrices, our definition focuses on the distinction between different quantum operations. In Section \ref{application}, we elaborate on the crucial role of this distinction in quantum algorithms. But in this section, we start with reinterpreting famous examples using our definition.

In Shor's algorithm \cite{shor1994algorithms} for finding the prime factors $p$ and $q$ of an integer $N$, a random integer $1<a<N$ is selected, and the quantum circuit is designed to find the period $r$ such that $a^r=1 \mod N$. The operators of quantum modular exponentiation (QME), denoted as $Ua^{2^i}$ for $0\leq i < 2n$ with $n=\lceil\log_2 N\rceil$, transform any $\ket{x}$ into $\ket{x\times a^{2^i} \mod N}$. For any $i$, if we apply $r$ operators $Ua^{2^i}$ one after the other, we will always transform $\ket{x}$ back to itself. Additionally, although it is useless for applications, any $Ua^{2^i}$ can be generated with $2^i$ operators $Ua^{2^0}$. Combining these two pieces of information, we can see that $Ua^{2^i}$ are elements of the quantum representation of $C_r$ with one generator $Ua^{2^0}$. Here, we re-interpret Shor's algorithm with our definition of quantum representation, which is unsurprising because Shor's inspiration came from Simon's algorithm to solve the hidden subgroup problem (defined in the following) when $G$ is abelian. When the input state of QME is $\ket{1}$ as demanded by the algorithm, after applying all the $Ua^{2^i}$ operators controlled by the uniform superposition, the output of the QME register becomes the superposition (usually not uniform) of $\ket{a^i \mod N}$ for $0\leq i < r$, which is the superposition of $C_r$-torsor. In particular, $C_r$-torsor is a subset of $C_{\varphi\left(N\right)}$-torsor, where $\varphi$ is the Euler's totient function and $\varphi\left(N\right)=\left(p-1\right)\times\left(q-1\right)$.

In Grover's algorithm \cite{grover1996fast} for searching an unstructured database, a uniform superposition over all states $\frac{1}{\sqrt{N}}\sum_{i=0}^{N-1}\ket{i}$ is first created. During each iteration step, the state vector is rotated with an angle of $\theta=2\arcsin{\frac{1}{\sqrt{N}}}$ on the $2$-dimensional subspace spanned by the initial superposition and the final target state that encodes the answer. Since the operator of each Grover's step is identical, Grover's algorithm forms a quantum representation of cyclic groups. When $N=2$, Grover's steps are isomorphic to $C_4$; and when $N=4$, they are isomorphic to $C_6$. Otherwise, Grover's steps never return the quantum state to the initial superposition, and they are all isomorphic to the same infinite cyclic group, $C_{\infty}$.

\section{Group to Quantum Circuits}\label{group to quantum circuits}

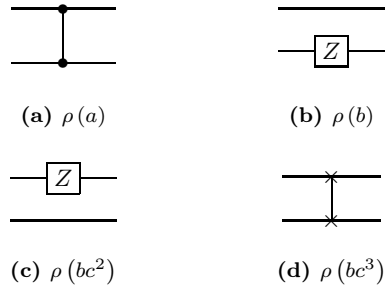
This section presents two approaches to transforming an arbitrary finite group $G$ into a set of quantum circuits. The first one, which can be described as "classic to quantum," involves constructing quantum circuits based on a classically computed complex representation of the group. The second one, which can be characterized as "quantum to classic," combines the group presentation with a variational quantum algorithm (VQA), which uses a classical optimizer to train a parameterized quantum circuit \cite{cerezo2021variational}. A toy example of the group $C_2\times D_4$ with a numerical simulation using the Qibo library\cite{qibo_paper} is available on GitHub\cite{Github}.

\subsection{Unitary Decomposition}

For a finite group $G$, suppose we are given a faithful complex representation. We can use the method in Ref. \cite{zee2016group} to transform it into a unitary representation $\widetilde{\rho}$, such that for each $g \in G$, $\widetilde{\rho}\left(g\right)$ is an $l\times l$ matrix with complex entries. Then we can construct another unitary and faithful representation $\rho: G \longrightarrow U\left(d\right)$ with a higher dimension by directly summing it with other unitary representations $\sigma_i$ (not necessarily faithful)

\begin{equation}
\rho\left(g\right)=
\begin{pmatrix}
\begin{tabular}{c|c|c}
$\sigma_1\left(g\right)$ &  $0$ & $0$\\
\hline
$0$ & $\widetilde{\rho}\left(g\right)$ & $0$ \\
\hline
$0$ & $0$ & $\sigma_2\left(g\right)$
\end{tabular}
\end{pmatrix}.
\end{equation}

In the proof of Theorem \ref{th}, $\rho$ is the mapping from $G$ to the permutation matrices of dimension $d$, where no extra $\sigma_i$ is required. When translated into quantum circuits, this type of operator provides an advantage by physically implementing $n$-control-qubit \textit{Toffoli} gates \cite{monz2009realization}.

The trivial representation can be chosen as $\sigma_i$ for simplicity, as in Eq. (\ref{eq_Ug}), where each element $g \in G$ is mapped into $I_k$. The dimension $k$ can be a number adapting to the quantum system, such as $k=d-l$.

\begin{equation}
\rho\left(g\right)=
\begin{pmatrix}
\begin{tabular}{c|c}
$\widetilde{\rho}\left(g\right)$ & $0$ \\
\hline
$0$ & $I_k$
\end{tabular}
\end{pmatrix}
\label{eq_Ug}
\end{equation}

The extra advantage of using the identity matrix to enlarge the representation is that it eliminates the effect of the global phase. In the faithful representation $\widetilde{\rho}$, every element of $G$ is mapped into a different matrix. On the other hand, in the trivial representation, all elements are mapped into $I_k$. Therefore, in the representation $\rho$, the identity element $e$ is the only element mapped into $I_{d}$. It does not exist an element $g$ in $G$ that verifies $\rho\left(g\right)=\exp\left({i\theta}\right)I_{d}$ with $\theta\neq 0$ because one block of the full matrix must be maintained as $I_k$. 

When $d=2^n$, after each $g\in G$ is mapped into a different $2^n\times 2^n$ unitary matrix, we can use the method provided in Ref. \cite{vartiainen2004efficient,li2013decomposition,krol2022efficient} to map it into a quantum circuit with a number of \textit{CNOT} gates $\mathcal{O}\left(4^n\right)$. In particular, any cyclic group can be mapped into single qubit gates. This is a straightforward construction in which the quantum advantage is not evident. However, because we are offering a general framework that includes arbitrary finite groups with any faithful complex representation, proof of advantage at this stage will be impossible unless constraints are added.

To give an order of magnitude, Monster group \cite{conway1985atlas}, the largest sporadic simple group $\mathbb{M}$, which contains $\sim 8\times 10^{53}$ elements, has $12$ genrators and a faithful representation of dimension $196884 < 2^{18}$. Therefore, $\mathbb{M}$ can be represented by $18$-qubit gates, where the $\mathbb{M}$-torsor is made of $18$-qubit states. This gives us a new perspective; besides a concept that appears during the study of quantum computing, QRFG might exist as a natural phenomenon, which is experimentally observable and non-artificial, that grants mystical finite groups a physical meaning. Otherwise, why do these structures emerge from the axioms?

\subsection{Variational Quantum Algorithm}

In the second method, we directly construct a quantum representation from the presentation of the group. Finite groups can be defined by their absolute presentations, which list the essential relations and irrelations that generators satisfy

\begin{equation}
G=\langle \underbrace{S}_\text{set of generators} | \underbrace{R}_\text{set of relations}, \underbrace{I}_\text{set of irrelations}\rangle.
\end{equation}

Every relation can be written in the formula of a word equal to the identity. A word in a group is defined as the product of group elements. If $G$ is generated by elements $g_1$ and $g_2$ that satisfy the relation $g_1 g_2=g_2 g_1$, it can be transformed into $g_2^{-1}g_1 g_2 g_1^{-1}=e$, where $g_2^{-1}g_1 g_2 g_1^{-1}$ is a word. The presentation of the cyclic group $C_8$ is $\langle g | g^8=e\rangle$. However, $g=e$ also satisfies the condition $g^8=e$. To eliminate confusion, we specify that certain words should not equal the identity. These are called irrelations. For $C_8$, the irrelation is $g^4\neq e$, the absolute presentation is $\langle g | g^8=e, g^4\neq e\rangle$.

The second method revolves around the central principle of formulating a quantum operator for each generator of the group. The key aspect is that applying these operators in the order dictated by the relations of the group preserves the quantum state of the system. In other words, there are no observable changes to any quantum state when the operators are applied according to the group's relations. However, when these operators are applied in the order of irrelations, noticeable changes occur in certain quantum states.

Then, a VQA can be devised to construct a quantum circuit for each generator, with the circuits depending on a set of classical parameters. These parameters can be adjusted iteratively within a quantum-classical optimization loop, aiming to minimize a predefined cost function.

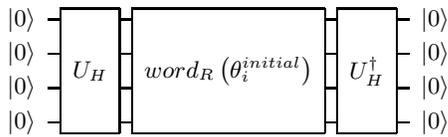
\begin{figure}[t]
\centering
\[
\begin{array}{c}
\Qcircuit @C=0.5em @R=0.5em
{
 \lstick{\ket{0}} & \multigate{3}{U_H} & \multigate{3}{word_{R}\left(\theta_i^{initial}\right)}  &\multigate{3}{U^\dagger_H} & \rstick{\ket{0}} \qw \\
 \lstick{\ket{0}} & \ghost{U_H} & \ghost{word_{R}\left(\theta_i^{initial}\right)} & \ghost{U^\dagger_H} & \rstick{\ket{0}} \qw\\
 \lstick{\ket{0}} & \ghost{U_H} & \ghost{word_{R}\left(\theta_i^{initial}\right)} & \ghost{U^\dagger_H} & \rstick{\ket{0}} \qw\\
 \lstick{\ket{0}} & \ghost{U_H} & \ghost{word_{R}\left(\theta_i^{initial}\right)} & \ghost{U^\dagger_H} & \rstick{\ket{0}} \qw
}
\end{array}
\]
\caption{A variational circuit to train a relation into identity. $U_H$ is a random unitary. The circuit is well-trained when it transform $\ket{0...0}$ into $\ket{0...0}$ with any $U_H$.}
\label{fig_relation}
\end{figure}

First, each generator is described as a variational ansatz with the same set of parameters. Here, we highlight that the ansatz for each generator can be set differently depending on the group. Each word is a concatenation of these ansatzes. 
Multiple circuits, as illustrated in FIG. \ref{fig_relation}, are simultaneously trained to maximize the amplitude of state $\ket{0...0}$ while minimizing the amplitudes of other states. Notably, only one quantum system of dimension $d$ is required, as the classical optimization of each loop is performed after the quantum state measurement. Besides the ground state of a qubit, $\ket{0}$ can also be the ground state of qutrit or qudit. This VQA works on a mixed-radix quantum architecture. A maximization (minimization) always exists if the ansatzes are properly designed. In the worst-case scenario, the trivial representation is obtained, where every generator is mapped into the identity circuit. 

In particular, some generators can first be implemented with unitary decomposition if it is more convenient. These two methods can be dynamically combined to reduce the parameter space. Similar parameters can be considered the same for the next training, and the circuits can be trained iteratively. 

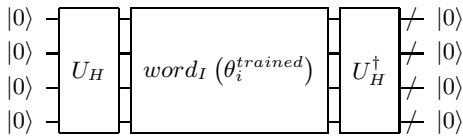
\begin{figure}[t]
\centering
\[
\begin{array}{c}
\Qcircuit @C=0.5em @R=0.5em
{
 \lstick{\ket{0}} & \multigate{3}{U_H} & \multigate{3}{word_{I}\left(\theta_i^{trained}\right)}  &\multigate{3}{U^\dagger_H} & {/} \qw & \rstick{\ket{0}} \qw \\
 \lstick{\ket{0}} & \ghost{U_H} & \ghost{word_{I}\left(\theta_i^{trained}\right)} & \ghost{U^\dagger_H} & {/} \qw & \rstick{\ket{0}} \qw\\
 \lstick{\ket{0}} & \ghost{U_H} & \ghost{word_{I}\left(\theta_i^{trained}\right)} & \ghost{U^\dagger_H} & {/} \qw & \rstick{\ket{0}} \qw\\
 \lstick{\ket{0}} & \ghost{U_H} & \ghost{word_{I}\left(\theta_i^{trained}\right)} & \ghost{U^\dagger_H} & {/} \qw & \rstick{\ket{0}} \qw
}
\end{array}
\]
\caption{A circuit to verify that an irrelation is not the identity. $U_H$ is a random unitary. The circuit is validated if there is some $U_H$ that transformed $\ket{0...0}$ into non $\ket{0...0}$.}
\label{fig_irrelation}
\end{figure}

After training, the quantum circuit for each generator can be reconstructed from the numerical output. Finally, the trained parameters are inserted into the verification circuits, as shown in FIG. \ref{fig_irrelation} to check that the word in the irrelation does not equal the identity, such that the quantum representation is faithful. With a large enough $d$ and universal ansatzes, the existence of a trained result validated by the verification circuits is justified by the Theorem \ref{th}. Furthermore, if the ansatzes are chosen specifically to avoid the effect of the global phase, we can obtain a classical and analytical representation of $G$ from the trained parameters. 

This method, like any other VQA, can be stuck in a "barren plateau" which has an exponentially small probability of escaping \cite{mcclean2018barren}. However, we will not elaborate on this limitation because it is impossible to study the technical details for such a general construction that works on arbitrary finite groups and can make use of any variational ansatz.

\section{Application}\label{application}

\subsection{Hidden Subgroup Problem}

The transformation of a finite group $G$ into quantum circuits holds practical significance, particularly for the gate-level implementation of quantum algorithms aimed at solving problems involving finite groups. For example, the hidden subgroup problem \cite{ettinger2004quantum}, a foundational topic in theoretical computer science.

\begin{definition}{Hidden subgroup problem (HSP)}

Let $G$ be a group, $X$ be a finite set, and an oracle function $f: G\rightarrow X$ that hides a subgroup $H\leq G$, which means for $g_1,g_2\in G$, $f\left(g_1\right)=f\left(g_2\right)$ if and only if $g_1$ and $g_2$ are on the same coset of $H$. On a quantum computer, $f$ is given as a unitary operator $O_f$ that is considered known. The task is to output the generating set of $H$.
\end{definition}

Shor's algorithm is a simple case of HSP, where $G=C_{\infty}$ and $H=C_r$, and $f$ transforms any integer $x$ into $a^x \mod N$. We highlight that $G$ is not $C_{2^{2n}}$ as the initial superposition on the register that controls QME. Because $r$ does not always devise $2^{2n}$, and according to Lagrange's theorem, $C_r$ can not be a subgroup of $C_{2^{2n}}$. However, only a finite number of elements can be encoded in superposition and the algorithm still works because it is a simple case.

The general quantum algorithm solves the HSP for a finite $G$ with polynomial (in $\log{|G|}$) query complexity, which is exponentially better than the best classical algorithm. But it requires exponential time, as in the classical case. In Section \ref{group to quantum circuits}, the two methods that we provided both involve an exponential time complexity as the upper bound, whether in circuit depth or training time. This drawback is consistent with the nature of HSP.

In this article, we do not discuss the principle of the algorithm and focus on the application of the unitary $O_f$ and the preparation of the initial state $\frac{1}{|G|}\sum_{g\in G}\ket{g}\ket{f\left(g\right)}$.

Multiple initial states are requested as inputs. The operator $O_f$ transforms $\ket{g}\ket{0...0}$ to $\ket{g}\ket{f\left(g\right)}$ for all $g\in G$. Therefore, before applying $O_f$, we need to prepare $\frac{1}{|G|}\sum_{g\in G}\ket{g}$, which is not indicated in the article, probably because it is considered trivial. However, a small attempt can reveal the technical difficulties behind the physical implementation. For example, encoding each element $\ket{g}$ in a computational basis is an easy solution. However, for a given $O_f$, two groups sharing the same number of elements cannot be distinguished.

In practice, preparing group elements as states actually means preparing the orbit of QRFG. The oracle operator $O_f$ should be implemented with the quantum representation of $G$ and $H$ because it controls on the register of $\ket{g}$ and targets on the register of $\ket{f\left(g\right)}$, to create the superposition of $G$-torsor and $H$-torsor. This operation grants an unavoidable necessity of faithfulness in our definition: without this property, the distinction between $G$ and its subgroup is not ensured to be physically observable. Moreover, Theorem \ref{th} provides proof that once the quantum algorithm that solves a finite group problem can be described at the abstract level, it can be performed at the gate level using various methods that work on different architectures.

\subsection{Quantum Finite Automata}

QRFG guarantees a finite number of accessible states with limited operators and can serve as the necessary building blocks for constructing quantum finite automata, which, as discussed by Moore and Crutchfield \cite{moore2000quantum}, harnesses the power of quantum mechanics to perform computation on finite languages \cite{say2014quantum,young2014quantum,kondacs1997power,lukac2009quantum}. The study of quantum languages for such machines remains an active area of research, here we take a popular definition given in Ref. \cite{say2014quantum}.

\begin{definition}{Quantum finite automata (QFA)}

A $m$-state Quantum finite automata is a five-tuple $M=\{Q,\Sigma,\{\mathcal{E}_{\sigma}|\sigma \in \Sigma\},q_1,Q_a\}$, with:

1) $Q=\{q_1,...,q_m\}$ is the set of states, $q_1 \in Q$ is the initial state, and $Q_a\subseteq Q$ is the set of accepting states.

2) $\Sigma$ is the alphabet. The set of all strings in the alphabet $\Sigma$ is $\Sigma^*$.

3) $\mathcal{E}_{\sigma}$ is a superoperator defined for each $\sigma\in\Sigma$, corresponding to a unitary operator $U_{\sigma}$.

Let $\omega\in\Sigma^*$ be the input. The computation starts in state $\rho_0=\ket{q_1}\bra{q_1}$. After reading each symbol, the defined superoperator is applied,

\begin{equation}
\rho_t=\mathcal{E}_{\omega_t}\left(\rho_{t-1}\right)=U_{\omega_t}\rho_{t-1}U^\dagger_{\omega_t},
\end{equation}
where $1 \leq t\leq |\omega|$. The $j$-th diagonal entry of $\rho$ represents the probability of the system being observed in the $q_j$ state. After reading the entire input, a measurement in the computational basis is made, and the input is accepted if one of the accepting states is observed. The overall accepting probability can be calculated as:

\begin{equation}
f_M\left(\omega\right)=\sum_{q_j\in Q_a} \rho_{j,j}.
\end{equation}

\end{definition}

The key to implementing QFA is to guarantee that, given an initial state $\ket{q_1}$, only a finite number of quantum states $\ket{q_j}$ can be accessed through any concatenation of $U_{\sigma}$. This is easy to realize with a finite group with $|G|=m$, where $U_\sigma$ forms its quantum representation and some vectors $\ket{q_j}$ in the computational basis forms its orbit. The structure of the group is not important. An explicit finite group $G$ can be chosen according to the given task and the experimental condition. For example, choose a $G$ such that $U_\sigma$ can be prepared with an easier operation, which under the circuit model means a shallower circuit. 

We can extend the definition, $\ket{q_1}$ does not have to be $\ket{0...0}$, and $\ket{q_j}$ does not have to be a state of the computational basis. Every quantum state in the orbit is required to be distinguishable with a minimum number of measurements, not necessarily one. This is achievable by maintaining an appropriate "distance" between them, defined with the Fubini–Study metric.

Now we can consider another perspective, and try to design an implementation of QFA without QRFG. First, we need to select the quantum states $\ket{q_1},...,\ket{q_m}$. Then we need to design the operation that transforms a given state to a target state according to the given task, for example, $U_{\sigma,1,2}\ket{q_1}=\ket{q_2}$. It is trivial to see that $U_{\sigma,i,j}$ generates a group, which does not need to be finite to be useful. The problem is whether the intermediate states and the final state remain within $Q$ or a simple and controllable superposition of elements in $Q$, for every $\omega\in\Sigma^*$. Otherwise, it is not a QFA because it is not finite. Without QRFG, we need to exhaustively verify every possible outcome before deciding on a method of implementation. This justifies the importance of the QRFG.

\section{Conclusion}

By revisiting the group axioms, we can gain insight into why the quantum representation is defined in terms of operators rather than states. Even in the absence of quantum properties such as unitarity, reversible operators still satisfy group axioms. Applying two operators in succession results in a new operator, and this combination of operators adheres to the associative property. There exists a unique identity element, representing the concept of doing nothing or having no effect. Applying an operator after its inverse leaves the system unchanged. 

In contrast, combining two quantum states to form a new state lacks a natural binary operation. Without the introduction of an operator, it is not possible to combine states meaningfully within the framework of group theory. It is through the quantum representation, which acts on these states, that the group structure can be defined and studied. Therefore, the choice to define quantum representation as operators rather than states is motivated by the fundamental properties of groups. Although states are important in quantum mechanics, it is the operators that enable the mathematical formulation and exploration of group theory within the context of quantum systems.

In this article, we present a formal definition of QRFG, employing concepts from both group theory and differential geometry. Our research establishes the existence of a quantum representation for any finite group, offering a solid theoretical insight. We propose two distinct methods for translating the generators of a group into quantum circuits, utilizing the gate decomposition of unitary matrices and VQA. To support our findings, we provide numerical simulations of a specific example accessible on an open-access platform. 

Moreover, we emphasize the practical significance of QRFG by demonstrating its application in the implementation of the quantum algorithm that solves HSP and in QFA. This reveals the potential and usefulness of our approach in addressing fundamental challenges in theoretical computer science. Overall, our work contributes to the understanding and advancement of the QRFG, offering theoretical foundations, practical implementation methods, and supporting numerical simulations.

\begin{acknowledgements}

The author extends their gratitude to S. Efthymiou for valuable support in Qibo, as well as to S. Li, S. Ramos-Calderer, I. Roth, Prof. G. Sierra, and Prof. J. I. Latorre for their helpful discussions. 

\end{acknowledgements}

%\bibliographystyle{unsrt}
%\bibliography{cite}

\end{document}